\documentclass[aps,pre,reprint]{revtex4-2}
\usepackage{amsfonts,amssymb,amsmath,bm}
\usepackage[dvips]{graphicx}
\usepackage{color}
\usepackage[dvipsnames]{xcolor}
\usepackage{hyperref}
\usepackage{times}
\usepackage{rotating}

\renewcommand{\Re}{\mathrm{Re}}

\begin{document}

\title{Profile of a Two-Dimensional Vortex Condensate Beyond the Universal Limit}

\author{Vladimir Parfenyev}\email{parfenius@gmail.com}

\affiliation{Landau Institute for Theoretical Physics, Russian Academy of Sciences, 1-A Akademika Semenova av., 142432 Chernogolovka, Russia}
\affiliation{National Research University Higher School of Economics, Faculty of Physics, Myasnitskaya 20, 101000 Moscow, Russia}

\date{\today}

\begin{abstract}
It is well known that an inverse turbulent cascade in a finite ($2 \pi \times 2 \pi$) two-dimensional periodic domain leads to the emergence of a system-sized coherent vortex dipole. We report a numerical hyperviscous study of the spatial vorticity profile inside one of the vortices. The exciting force was shortly correlated in time, random in space, and had a correlation length $l_f = 2\pi/k_f$ with $k_f$ ranging from $100$ to $12.5$. Previously, it was found that in the asymptotic limit of small-scale forcing, the vorticity exhibits the power-law behavior $\Omega(r) = (3 \epsilon/\alpha)^{1/2} r^{-1}$, where $r$ is the distance to the vortex center, $\alpha$ is the bottom friction coefficient, and $\epsilon$ is the inverse energy flux. Now we show that for a spatially homogeneous forcing with finite $k_f$ the vorticity profile becomes steeper, with the difference increasing with the pumping scale but decreasing with the Reynolds number at the forcing scale. Qualitatively, this behaviour is related to a decrease in the effective pumping of the coherent vortex with distance from its center. To support this statement, we perform an additional simulation with spatially localized forcing, in which the effective pumping of the coherent vortex, on the contrary, increases with $r$ and show for the first time that in this case the vorticity profile can be flatter than the asymptotic limit.
\end{abstract}

\maketitle

\section{Introduction}

The 2D Navier-Stokes equation forced at intermediate scales favors the transfer of energy to larger scales, a phenomenon known as an inverse cascade~\cite{kraichnan1967inertial, leith1968diffusion, batchelor1969computation}. Already the first experiments~\cite{sommeria1986experimental, paret1998intermittency} and numerical simulations~\cite{smith1993bose, smithr1994finite, borue1994inverse} on two-dimensional turbulence showed that in a finite domain with low bottom friction, the inverse cascade leads to the accumulation of energy at the system size and formation of coherent vortex structures. Subsequent numerical~\cite{chertkov2007dynamics, laurie2014universal, frishman2018turbulence} and experimental~\cite{xia2009spectrally,orlov2018large} studies demonstrated that these vortices have well-defined isotropic mean profiles with a radial power-law decay of vorticity in the inner region. Analytical progress can be made if the condensate is strong compared to turbulence. Then the self-action of turbulent pulsations is small compared to the action of the mean flow, and a quasi-linear passive theory can be developed~\cite{kolokolov2016structure, kolokolov2016velocity, frishman2017culmination}. If turbulence is excited at asymptotically small scales, the treatment allows one to derive an explicit formula for the condensate vorticity profile, $\Omega(r) = (3 \epsilon/\alpha)^{1/2} r^{-1}$, where $\alpha$ is the bottom friction coefficient and $\epsilon$ is the inverse energy flux. Since the profile does not depend on the type of small-scale dissipation and small-scale forcing, it was called universal~\cite{laurie2014universal}. Note that a similar quasi-linear approach was also used to describe jets in rectangular periodic domains~\cite{frishman2017jets}, on a sphere~\cite{falkovich2016interaction}, and on a periodic beta plane~\cite{woillez2017theoretical, woillez2019barotropic}.

Theoretical predictions in the universal limit agree with the results of numerical studies~\cite{laurie2014universal, frishman2018turbulence}. In these works, the simulations were carried out in periodic domain with dimensions of $2 \pi \times 2 \pi$, and the forcing wave number was at least $k_f = 100$. In laboratory experiments, the length of the inertial interval for the inverse cascade is much shorter, so the question arises of how the vorticity profile will change as $k_f$ decreases. In earlier experiments, the dependence $\Omega(r) \propto r^{-1.25}$ was reported, although a rather large error in the determination of the exponent should be noted~\cite{xia2009spectrally}. Here we systematically study the issue and demonstrate that for a random forcing with spatially homogeneous statistics the vorticity profile becomes steeper than the universal limit, with the difference increasing with the pumping scale but decreasing with the Reynolds number at the forcing scale. We discuss our results in the context of the Reynolds equation for the mean polar velocity of the vortex and conclude that the main distinction from the universal limit is that the effective pumping of the coherent vortex falls off as $r$ increases, leading to the observed steeper behaviour. Comparison with the quasi-linear passive theory~\cite{kolokolov2016structure} shows that it overestimates the effective pumping of the coherent vortex, and we suppose that this is caused by the self-action of fluctuations, which we model by the eddy viscosity resulting in a reasonable agreement with the simulations.

Next, the following question naturally arises: is the self-organization of coherent vortices with a vorticity profile flatter than the universal limit possible? We show that the answer is positive, and to demonstrate this we consider the forcing that has been spatially localized at two small spots. It turns out that the external forcing has no pinning effect on the coherent vortices and most of the time they are far from the pumped regions. In this case, the effective pumping of vortices mostly occurs due to velocity fluctuations that they meet with their outer edges during their movement through the system. Therefore, the effective pumping of vortex has a maximum at a certain distance from its center and decreases as one moves towards the center of the vortex. This effective pumping profile results in a vorticity distribution that is flatter than the universal limit. 

\begin{figure*}[t]
\centering{\includegraphics[width=\linewidth]{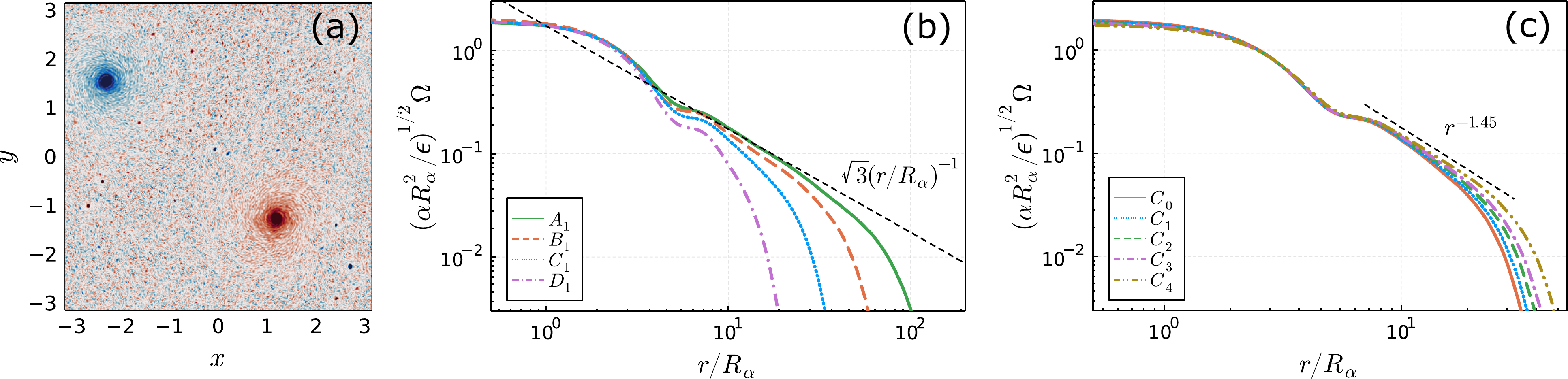}}
\caption{The stationary state for spatially uniform pumping: (a) snapshot of the vorticity field ($512^2$) for DNS run $A_1$, (b,c) radial profile of the mean vorticity for DNS runs $A_1$-$D_1$ and $C_0$-$C_4$. The straight dashed line in Fig.~\ref{fig:1}b corresponds to the universal limit, $\Omega(r) = (3 \epsilon/\alpha)^{1/2} r^{-1}$.} 
\label{fig:1}
\end{figure*}

\section{Numerical Methods}

We solve the incompressible forced Navier-Stokes equation with hyperviscous dissipation and linear bottom friction for a fluid with unit density in 2D:
\begin{equation}\label{eq:1}
\partial_t \bm v + (\bm v \nabla) \bm v = - \nabla p -\alpha \bm v - \nu (-\nabla^2)^{q} \bm v + \bm f,
\end{equation}
where $\bm v$ is 2D velocity, $p$ is the pressure, $\alpha$ is the friction coefficient, $\nu$ is the hyperviscosity, and $\bm f$ is a random forcing. The domain is a doubly periodic square box of size $L=2 \pi$. We work with an isotropic, shortly correlated in time forcing acting in a narrow ring in Fourier space centered on wave number $k_f$, with $\varepsilon = \langle \bm v \cdot \bm f \rangle$ the average energy injection rate, and angular brackets denote time-averaging. In an unbounded system, the inverse energy cascade is terminated by the bottom friction at the scale $L_{\alpha} \sim \varepsilon^{1/2} \alpha^{-3/2}$, where a balance between the energy flux and the bottom friction is achieved~\cite{boffetta2012two}. We assume that the friction coefficient $\alpha$ is small enough, so that $L<L_{\alpha}$, and then the energy, transferred to the domain size $L$ by the inverse cascade, is accumulated there, giving rise to a mean (coherent) flow.

\begin{table}[b]
\begin{center}
\begin{tabular}{c|c|c|c|c|c}
\hline \hline
         run   &   grid  &  $k_f$ & $\nu$ & $\Re$ & $\epsilon$ \\ \hline
         $A_1$     &    512  &  100 & $5\cdot 10^{-35}$ & 304 & $1.94\cdot 10^{-4}$  \\
         $B_1$     &    512  &  50  & $2 \cdot 10^{-30}$ & 313 & $2.01 \cdot 10^{-4}$ \\
         $C_1$     &    256  &  25  & $8 \cdot 10^{-26}$ & 323 & $2.10 \cdot 10^{-4}$ \\
         $D_1$     &    256  &  12.5& $3.5 \cdot 10^{-21}$ & 305 & $2.17 \cdot 10^{-4}$ \\
         $C_0$ &   256   &  25 & $2.4 \cdot 10^{-25}$ & 108 & $1.88\cdot 10^{-4}$ \\
         $C_2$ &   256   &  25 & $2.4 \cdot 10^{-26}$ & 1078 & $2.30 \cdot 10^{-4}$ \\
         $C_3$ &   256   &  25 & $0.8 \cdot 10^{-26}$ & 3235 & $2.46 \cdot 10^{-4}$ \\
         $C_4$ &   256   &  25 & $0.8 \cdot 10^{-27}$ & 32348 & $2.73 \cdot 10^{-4}$ \\
     \hline \hline
\end{tabular}
\caption{Parameters for the DNS runs.}
\label{tab:1}
\end{center}
\end{table}

DNS results are obtained by integrating (\ref{eq:1}) in the vorticity formulation using the GeophysicalFlows.jl pseudospectral code~\cite{GeophysicalFlowsJOSS}, at resolution $256^2$ and $512^2$, with parameters $\varepsilon = 3.5 \cdot 10^{-4}$, $\alpha = 10^{-4}$, and $q=8$. The pumping covariance spectrum is Gaussian with mean $k_f$ and standard deviation $\delta_f=1.5 \ll k_f$. The high degree of hyperviscosity allows simulations to be performed with relatively low spatial resolution~\cite{chertkov2007dynamics, laurie2014universal, frishman2018turbulence, frishman2017jets}. The initial condition in all our simulations is a state of rest, and each simulation is run until the system reaches a non-equilibrium stationary state, observed by the saturation of the total kinetic energy. After that, the simulation continued for some time, necessary to collect statistics. The time step is fixed for each simulation and it satisfies $\Delta t < c_0 \Delta x/v_{max}$, where $\Delta x$ is the grid spacing, $v_{max}$ is the maximum value of the velocity field projections on the axes of the Cartesian coordinate system, and $c_0$ is equal to $0.3-0.5$. The time step $\Delta t$ is also the correlation time of the exciting force $\bm f$.

\begin{figure}[b]
\centering{\includegraphics[width=0.9\linewidth]{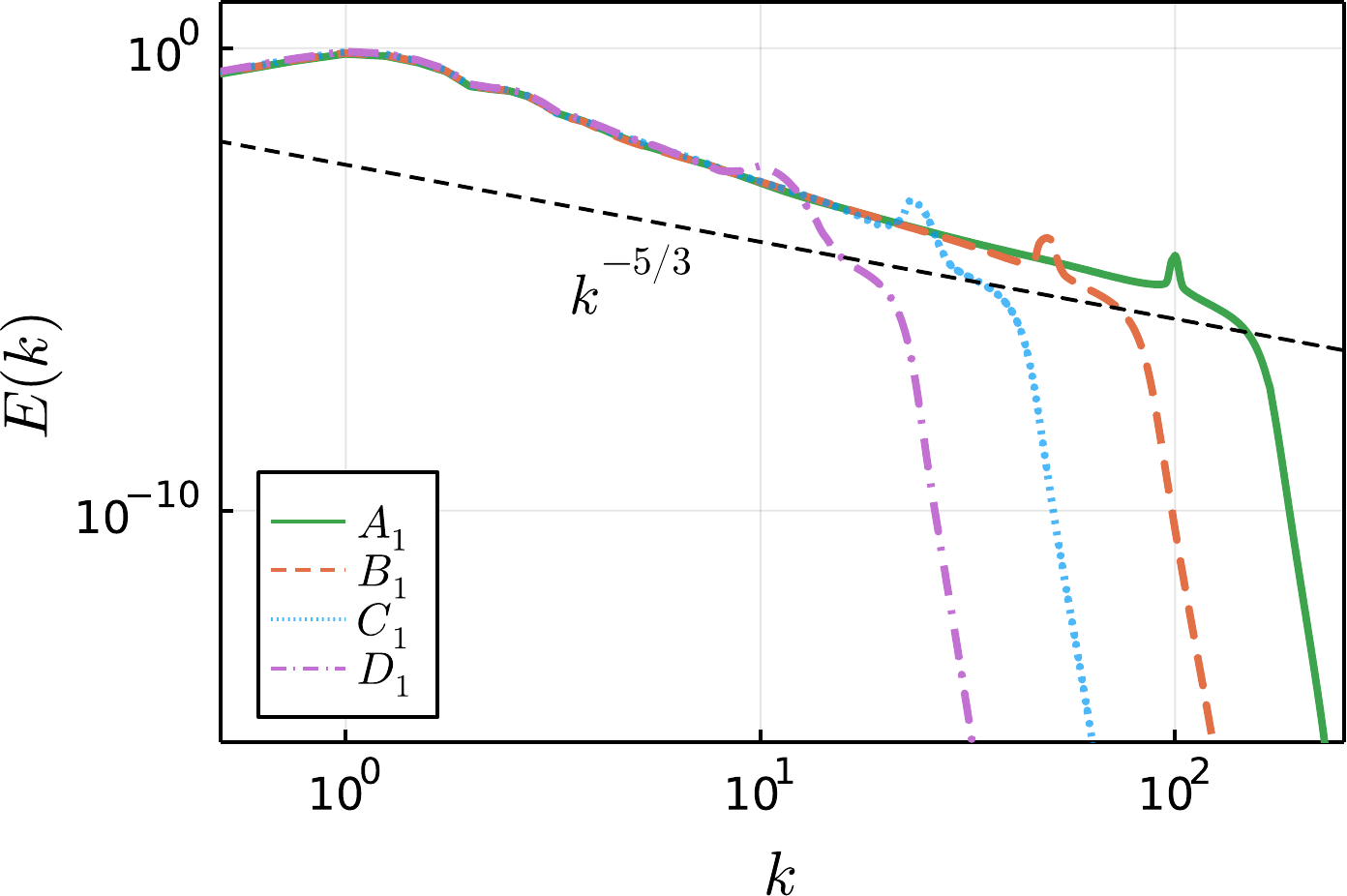}}
\caption{Energy spectra for DNS runs $A_1$-$D_1$.}
\label{fig:2}
\end{figure}

In the first set of simulations ($A_1$-$D_1$), we vary the pumping wave number in the range from $k_f=100$ to $12.5$, while adjusting the hyperviscosity $\nu$ so that the Reynolds number at the forcing scale $\Re = \varepsilon^{1/3}/(\nu k_f^{46/3})$ is about 300, see Table~\ref{tab:1}. In agreement with previous studies~\cite{laurie2014universal,frishman2017jets,frishman2018turbulence}, the flow reaches a condensate steady state, taking the form of a system-sized vortex dipole. The vortices drift slowly over time, with fast turbulent pulsations superimposed onto them (see Fig.~\ref{fig:1}a and video~\cite{SM}). We find that in all cases a significant part of the injected energy $\varepsilon$ is dissipated due to hyperviscosity at high wave numbers $k>k_f$. This is consistent with recent numerical studies~\cite{frishman2017jets,frishman2018turbulence}, but differs from the theoretical analysis of the universal limit~\cite{kolokolov2016structure, frishman2017culmination}, where it is assumed that all the energy is dissipated by bottom friction on large scales. To estimate the inverse energy flux from numerical data, we compute the energy dissipation rate by bottom friction during the steady state regime, $\epsilon = \alpha \langle \int dxdy \, \bm v^2/L^2 \rangle$. The corresponding values are given in Table~\ref{tab:1}, and henceforce they are used to normalize velocities by $(\epsilon/\alpha)^{1/2}$. This estimate is justified because the main contribution to the total energy of the system comes from large scales corresponding to the coherent vortex dipole, see Fig.~\ref{fig:2}. The slopes of the spectra are steeper than $-5/3$ due to the presence of condensate, in agreement with previous studies~\cite{chertkov2007dynamics, frishman2017jets, chan2012dynamics}. The relative fluctuations of the total energy of the system are small and do not exceed $1\%$ even for the case with the largest forcing scale under consideration.

\begin{figure}[t]
\centering{\includegraphics[width=0.9\linewidth]{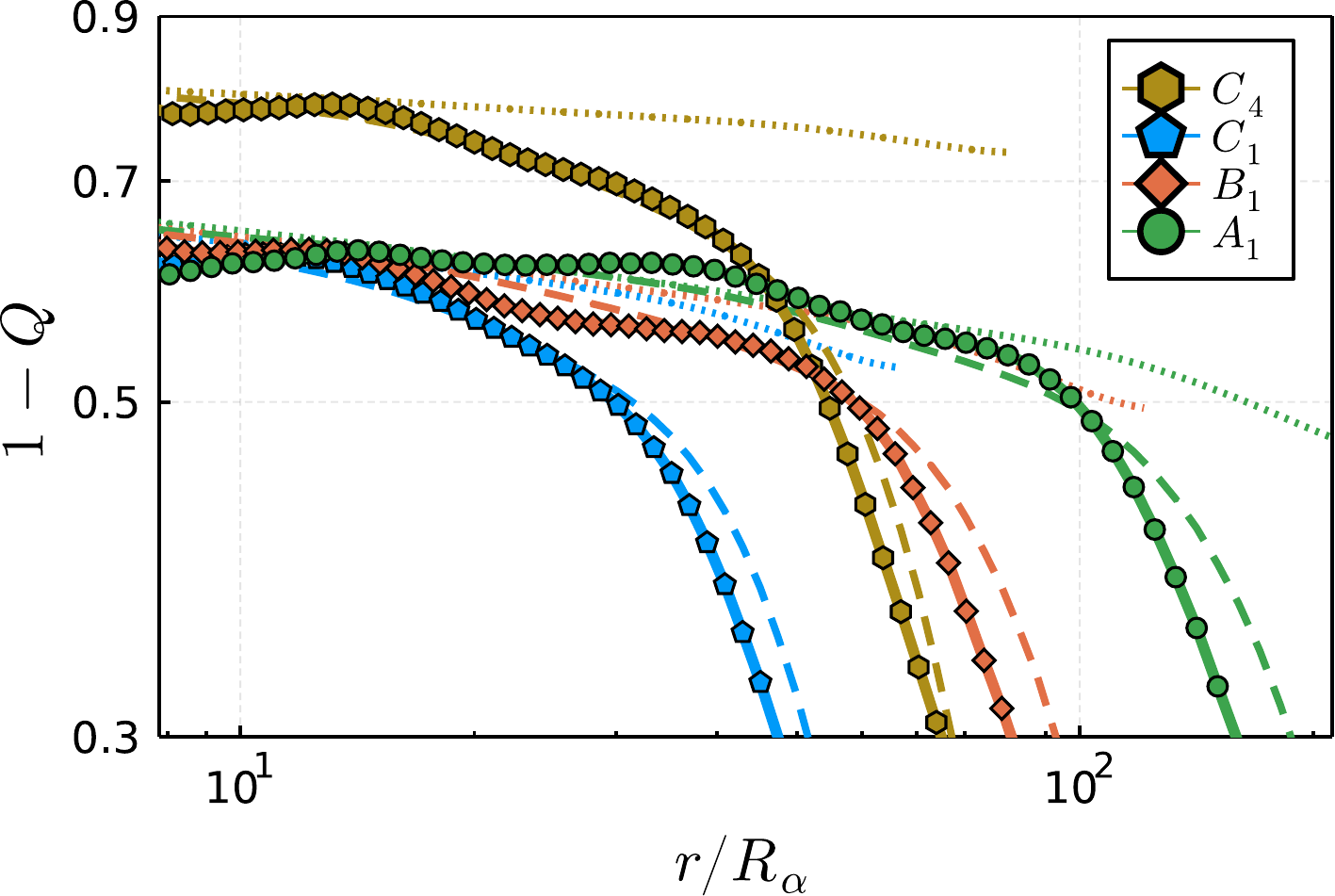}}
\caption{The dependence of $1-Q$ on the distance $r$ from the vortex center in log-log scale calculated by expression (\ref{eq:4}) (solid lines with markers), using the quasi-linear theory (\ref{eq:kolokolov}) (dotted lines), and using the quasi-linear theory with the turbulent diffusion correction (dashed lines).}
\label{fig:3}
\end{figure}

To obtain the mean vorticity distribution, we shift the individual snapshots describing the instantaneous vorticity field so that the center of one of the vortices (identified by the vorticity maximum) is always at the center of the domain, and then average over all snapshots corresponding to the non-equilibrium steady state. The resulting vorticity distribution $\Omega$ inside the vortex is highly isotropic, and it can be also described in terms of the mean polar velocity $U$ depending on the distance $r$ from the vortex center, $\Omega = (1/r) \partial_r (rU)$. The mean polar velocity satisfies the Reynolds equation~\cite{kolokolov2016structure}
\begin{equation}\label{eq:2}
\alpha U + \nu (-\nabla^{2})^{q} U = -\left( \partial_r + \dfrac{2}{r}\right) \langle u_r u_{\phi}  \rangle,
\end{equation}
where $u_r$ and $u_{\phi}$ are radial and polar components of the turbulent velocity fluctuations, and angular brackets denote time-averaging. Therefore, the coherent vortex maintains its existence due to the Reynolds shear stress $\langle u_r u_{\phi} \rangle$ (right-hand side), which balances the dissipative terms inside the vortex (left-hand side). The hyperviscous term is significant inside the vortex core~\cite{kolokolov2016structure}, at distances $r \lesssim R_{\alpha} = (\nu/\alpha)^{1/2q}$, and can be neglected outside, where the mean vorticity profile reveals the behaviour close to power-law for small-scale pumping, see Fig.~\ref{fig:1}b. Note that the velocity profile inside the viscous core was analyzed in detail in Refs.~\cite{parfenyev2021influence, doludenko2021coherent} in the case of an ordinary ($q=1$) viscous dissipation. This regime may be of interest for experiments with thin soap films~\cite{kellay2002two} or freely suspended smectic films~\cite{parfenyev2016nonlinear, yablonskii2017acoustic}, where friction against the bottom is absent. In what follows, we will focus on distances beyond the vortex core. We will use the scale $R_{\alpha}$ to normalize all distances.

Fig.~\ref{fig:1}b shows that the mean vorticity profile of run $A_1$ is close to the universal limit, but as the forcing scale increases (runs $B_1$-$D_1$), the vorticity profiles become steeper. To study the dependence of the profile slope on the Reynolds number (runs $C_0-C_4$), we fix the pumping wave number $k_f=25$ and change the hyperviscosity according to Table~\ref{tab:1}. The resulting mean vorticity profiles are presented in Fig.~\ref{fig:1}c. It can be concluded that an increase in the Reynolds number at the forcing scale results in flatter vorticity profiles.

\section{Effective Pumping of the Vortex}

\begin{figure}[t]
\centering{\includegraphics[width=0.9\linewidth]{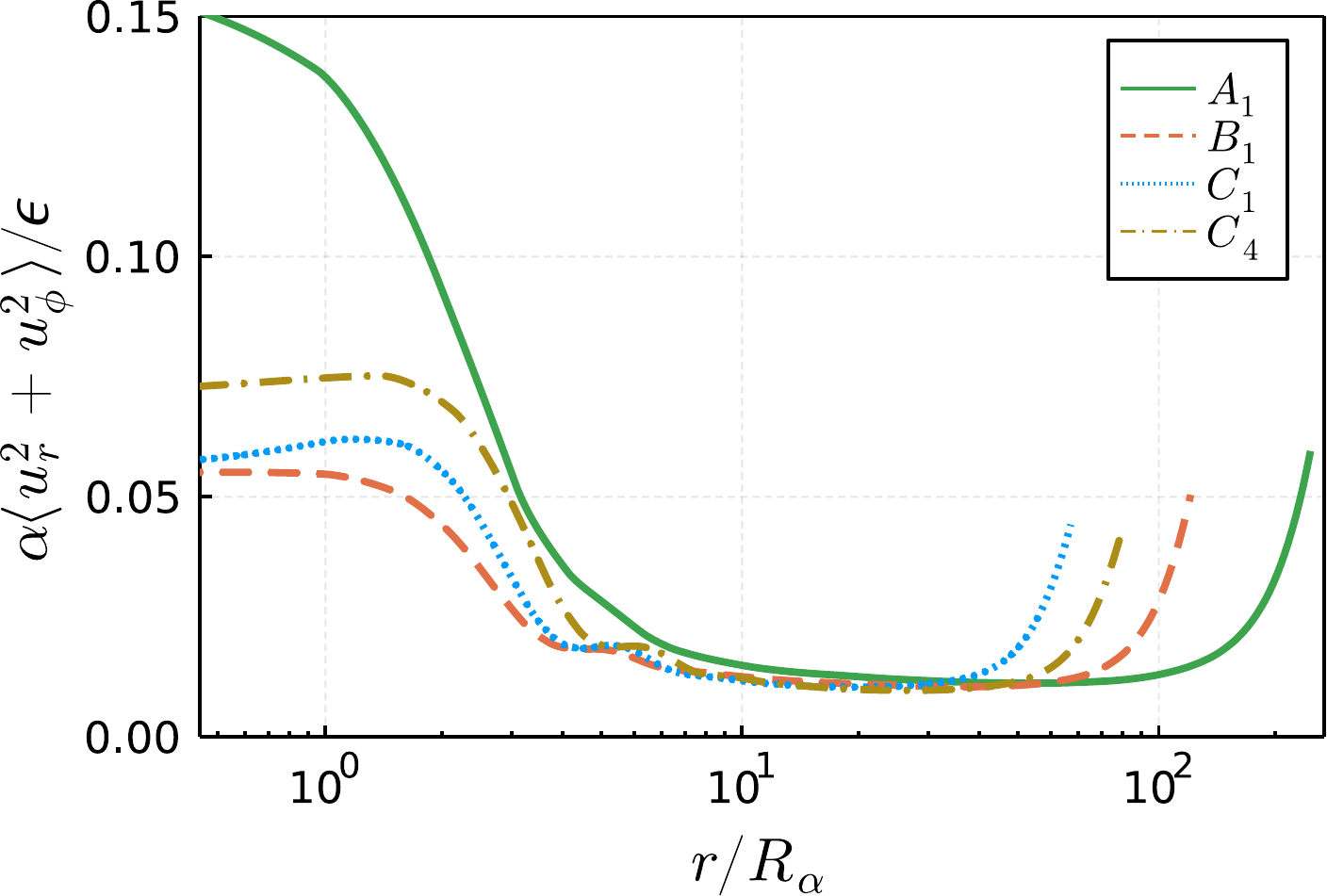}}
\caption{The intensity of velocity fluctuations as a function of the distance $r$ to the vortex center.}
\label{fig:4}
\end{figure}

To discuss the obtained results, let us analyze how and why they differ from the universal limit. Recall that the universal limit formally follows from Eq.~(\ref{eq:2}), if relation $\langle u_r u_{\phi} \rangle = \varepsilon(1-Q)/\Sigma$ is used for the Reynolds stress component, where $\Sigma = r \partial_r (U/r)$ describes the mean-flow shear rate, and if one neglects the hyperviscous term, since the region $r \gtrsim R_{\alpha}$ is analyzed~\cite{kolokolov2016structure}:
\begin{equation}\label{eq:3}
\alpha U = -\left( \partial_r + \dfrac{2}{r}\right) \dfrac{\varepsilon}{\Sigma} (1-Q).
\end{equation}
Here $Q$ can be thought of as a phenomenological parameter that describes the effective pumping intensity $\varepsilon (1-Q)$ of the coherent vortex. Now let us try to find the power-law solution of this equation in the form $\Omega \propto r^{-\beta}$, $U \propto r^{1-\beta}$, etc., and we will immediately obtain that the value of $1-Q$ should depend on $r$ as $1-Q \propto r^{2(1-\beta)}$. The universal limit corresponds to $\beta=1$ and then the solution of this equation is $U=(3 \epsilon/\alpha)^{1/2}$ and, accordingly, $\Omega = (3 \epsilon/\alpha)^{1/2} r^{-1}$, where $\epsilon=\varepsilon(1-Q)$. However, if $\beta>1$, then the effective pumping intensity of the coherent vortex $\varepsilon(1-Q)$ should decrease with increasing $r$. Can we support this statement quantitatively?

\begin{figure*}[t]
\centering{\includegraphics[width=\linewidth]{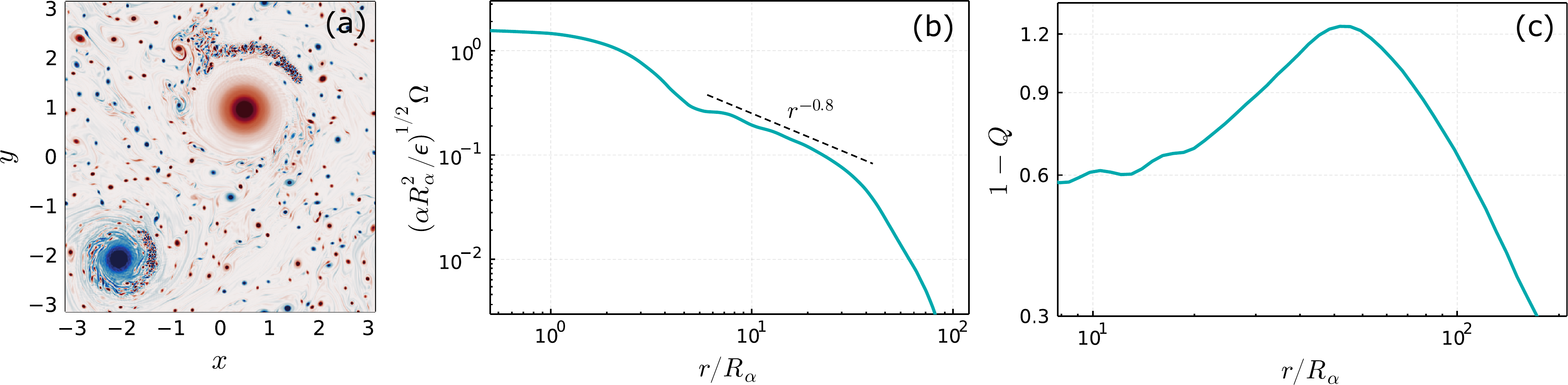}}
\caption{The stationary state for spatially localized pumping: (a) snapshot ($512^2$) of the vorticity field, (b) radial profile of the mean vorticity demonstrates behaviour flatter than the universal limit, (c) the dependence of $1-Q$ on the distance $r$ from the vortex center in log-log scale calculated by expression (\ref{eq:4}) based on DNS data.}
\label{fig:5}
\end{figure*}

Instead of power-law analysis of Eq.~(\ref{eq:3}), which is not well suited to our DNS runs because the intervals of linear behaviour on the log-log plots are quite short (see Fig.~\ref{fig:1}), we can integrate it directly
\begin{equation}\label{eq:4}
  1-Q = - \dfrac{\alpha \Sigma(r)}{\varepsilon r^2} \int_0^r d\xi \, \xi^2 U(\xi).
\end{equation}
From numerical simulations, we know the dependencies $U(r)$ and $\Sigma(r)$, and thus can calculate the dependence of $1-Q$ on $r$. Fig.~\ref{fig:3} shows the results for parameters corresponding to runs $A_1$-$C_1$ and $C_4$, which are chosen for illustrative purposes (solid lines with markers). One can conclude that the effective pumping intensity $\varepsilon(1-Q)$ of the coherent vortex actually decreases with increasing distance $r$ from its center, and the decrease is faster for steeper vorticity profiles. Note also that the effective pumping increases with increasing Reynolds number if the forcing scale is fixed.

The value of $Q$ can be also found theoretically in the framework of a quasi-linear treatment of relatively weak turbulent pulsations against the background of a strong coherent vortex (see Ref.~\cite{kolokolov2016structure} and Appendix~\ref{app:A})
\begin{eqnarray}\label{eq:kolokolov}
\nonumber
&Q \simeq \displaystyle 2 \nu \int_0^{\infty} d \tau \int \dfrac{d^2 \bm k \, k^2 \chi(\bm k) }{(2 \pi)^2} [(k_1 - \Sigma \tau k_2)^2 + k_2^2]^{q-1}& \\
&\displaystyle \times \exp \left[ -2 \int_{0}^{\tau} d \tau' \, \Gamma \left( \sqrt{(k_1 - \Sigma \tau' k_2)^2 + k_2^2} \right)  \right],&
\end{eqnarray}
where $\Gamma (k) = \alpha + \nu k^{2q}$ describes dissipation including both the bottom friction and the hyperviscosity term. For our parameters, the pumping covariance spectrum can be safely replaced by $\chi(\bm k) = 2 \pi \delta(k-k_f)/k_f$ and using the dependence $\Sigma(r)$ obtained from DNS, we can calculate $1-Q$, see dotted lines in Fig.~\ref{fig:3}. It turns out that the quasi-linear theory correctly describes the effective pumping of the coherent vortex far from the outer region, but overestimates it on the periphery, especially for relatively small values of $k_f$.

We suppose that this overestimation is due to the self-action of fluctuations, which was neglected in the quasi-linear theory. Fig.~\ref{fig:4} shows the intensity of velocity fluctuations, which weakly depends on the distance $r$ to the vortex center outside the viscous core. Based on these results, one can find that velocity fluctuations are indeed small compared to the mean flow, $\langle u_r^2 + u_{\phi}^2 \rangle^{1/2}/U \sim 0.1$. However, at the same time, the nonlinear
self-action of fluctuations is significant compared to the dissipative terms at the forcing scale, which determine the value of $Q$, i.e. $|(\bm u \nabla) \bm u| \gg |\Gamma (k_f) \bm u|$. To take into account the self-action of fluctuations phenomenologically, we propose to add an additional term $\nu_T k^2$ into $\Gamma(k)$, corresponding to eddy viscosity. Note that in this case there is also an additional contribution to the value of $Q$ proportional to the parameter $\nu_T$, see Appendix~\ref{app:B}. The correlation time of velocity fluctuations can be estimated as $1/|\Sigma(r)|$, and therefore the turbulent viscosity can be modelled as $\nu_T = \gamma/|\Sigma(r)|$, where $\gamma$ is a free parameter that is chosen to match the theory with the DNS. We found $\gamma \sim 4 \cdot 10^{-6}$ for run $A_1$, $\gamma \sim 2 \cdot 10^{-5} $ for run $B_1$, $\gamma \sim 10^{-4} $ for run $C_1$, and $\gamma \sim 1.5 \cdot 10^{-4}$ for run $C_4$. The results are shown in Fig.~\ref{fig:3} with dashed lines and they demonstrate a reasonable agreement with the simulations.

\section{Spatially Localized Forcing}

Finally, we would like to address the question --- is the self-organization of coherent vortices with a vorticity profile flatter than the universal limit possible? To the best of our knowledge, the observation of such vortices has not been reported in the literature until now. In accordance with the previous discussion, the existence of such vortices is feasible if the effective pumping intensity of the coherent vortex increases with distance from its center. To implement such situation, we consider a random forcing localized in space in two regions with radii $a=0.1$, small compared to the expected sizes of coherent vortices, but large compared to the pumping scale ($k_f=100$). The centers of the regions are located at points with coordinates $(\pm L/4, \pm L/4)$. In comparison with DNS run $A_1$, the amplitude of the external force was increased so that the power pumped into the system, averaged over the entire domain area, remained approximately the same.

\begin{figure}[t]
\centering{\includegraphics[width=0.8\linewidth]{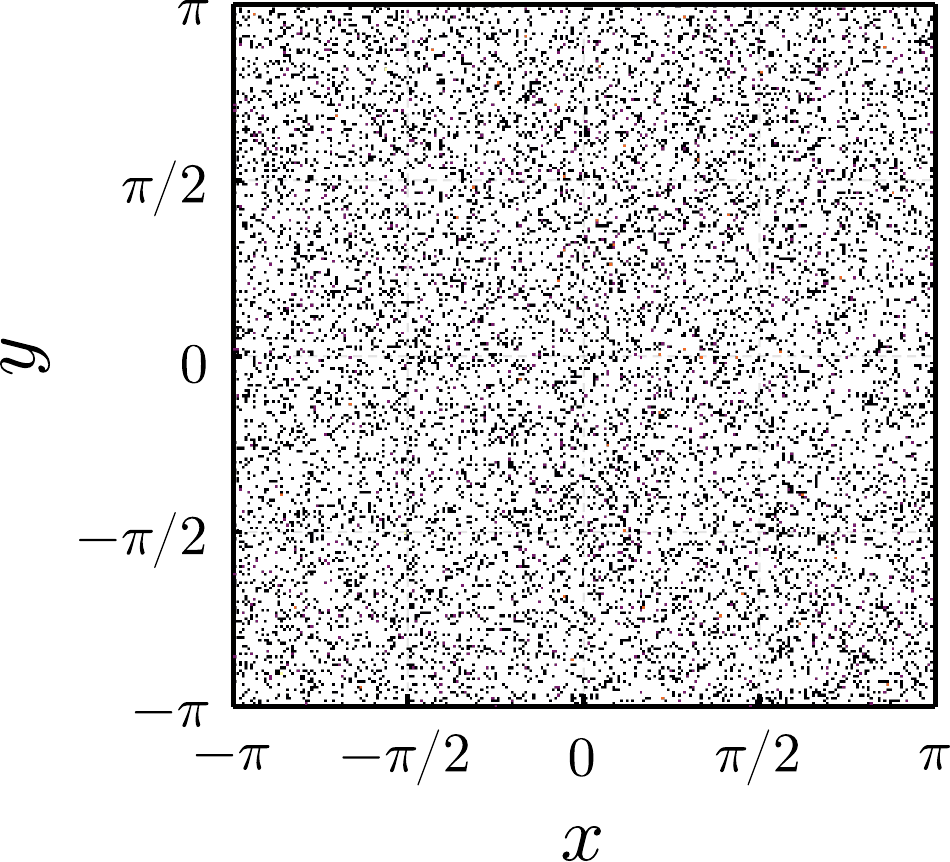}}
\caption{Each point corresponds to the position of the (positive) vortex center. About 12000 snapshots were analyzed.}
\label{fig:6}
\end{figure}

After some time, a pair of coherent vortices rotating in opposite directions is formed in the system, see Fig.~\ref{fig:5}a and video~\cite{SM}. In the stationary state, vortices slowly move through the system in random directions. The external forcing has no pinning effect on the vortices, see Fig.~\ref{fig:6}. So, the vortices explore all space uniformly.

The pumping creates fuel (turbulent pulsations) for the vortex. In the case of uniform pumping, at each moment of time, an external force excites fluctuations inside the vortex, which feed it. Now fluctuations are excited only in small regions, and the vortex spends most of the time away from them. In the process of its wandering, the vortex is fed by fluctuations that are encountered on its way. Thus, fluctuations penetrate into the vortex mostly from the outer region. In this case, it is reasonable to expect that the effective pumping of the vortex will have a maximum at some distance from its center. According to the previous analysis, such profile of the effective pumping should result in a vorticity distribution that is flatter than the universal limit.

The above qualitative reasoning can be supported by quantitative observations. Fig.~\ref{fig:5}b shows the radial profile of the mean vorticity distribution. It exhibits behavior flatter than the universal limit, in line with our expectations. Fig.~\ref{fig:5}c presents the dependence of the effective pumping intensity of the coherent vortex on distance from its center. The value of $1-Q$ increases with $r$ inside the vortex and has a maximum at some distance from the vortex center, in accordance with the above qualitative arguments.

\section{Conclusion}

To conclude, we performed hyperviscous numerical study of the vortex condensate in forced 2D turbulence. In the case of spatially uniform pumping, the vorticity profile is steeper than the universal limit, with the difference increasing with the pumping scale but decreasing with the Reynolds number at the forcing scale. An analysis of the Reynolds equation for the mean polar velocity of a coherent vortex led us to the conclusion that this behaviour corresponds to a decrease in the effective pumping intensity of the coherent vortex with distance from its center. Inspired by this observation, we performed an additional simulation with spatially localized forcing, in which the effective pumping intensity of the coherent vortex, on the contrary, increases with $r$ and show that in this case the vorticity profile becomes flatter than the universal limit. Our findings demonstrate that spatially inhomogeneous forcing opens up an additional degree of freedom for controlling the self-organization of coherent vortices.

To explore the possibilities of this additional degree of freedom in future studies, it is reasonable to replace the periodic boundary conditions with no-slip or stress-free walls. In such systems, the positions of coherent vortex structures are almost fixed due to geometric constraints~\cite{xia2009spectrally, molenaar2004angular, doludenko2021coherent, gallet2013two}, and therefore it will be relatively easy to design the desired spatial profile $\varepsilon(r)$ of pumping in the reference frame associated with the vortex.

Despite the relative simplicity of the system under study, the obtained results can be used to understand the processes of self-organization of large-scale vortex currents in the atmosphere and oceans. Since the pumping of such currents occurs in a non-uniform manner, further research in this direction looks promising. In addition to the mean velocity profile inside vortices, an important object of study is the velocity of a coherent vortex as a whole and its statistical properties. In the future, this will help to better understand the motion of hurricanes.

\acknowledgments

I wish to thank Vladimir Lebedev, Igor Kolokolov, and Sergey Vergeles for helpful discussions. The work was supported by the Russian Ministry of Science and Higher Education, project No. 075-15-2022-1099, and by the Foundation for the Advancement of Theoretical Physics and Mathematics ''BASIS''. Simulations were performed on the cluster of the Landau Institute.

\appendix

\section{Q-value calculation}\label{app:A}

The starting point of our analyses is expression (A12) in Ref.~[14]
\begin{widetext}
\begin{equation}
Q = 2 \int_0^{\infty} d \tau \int \dfrac{d^2 \bm p \, p^2 \chi(\bm p) }{(2 \pi)^2} \dfrac{\Gamma \left( \sqrt{(p_1 - \Sigma \tau p_2)^2 + p_2^2} \right)}{(p_1 - \Sigma \tau p_2)^2 + p_2^2} \exp \left[ -2 \int_{0}^{\tau} d \tau' \, \Gamma \left( \sqrt{(p_1 - \Sigma \tau' p_2)^2 + p_2^2} \right)  \right],
\end{equation}
where $\Gamma (k) = \alpha + \nu k^{2q}$ describes dissipation including both the bottom friction $\alpha$ and the hyperviscosity term, $\chi(\bm k)$ is the pumping covariance spectrum, and $\Sigma(r)$ is the mean-flow shear rate known from DNS. It is convenient to represent this expression as the sum of two terms, $Q = Q_{\alpha} + Q_{\nu}$, where
\begin{eqnarray}
&\displaystyle Q_{\alpha} = 2 \alpha \int_0^{\infty} d \tau \int \dfrac{d^2 \bm p}{(2 \pi)^2} \dfrac{\chi(\bm p) p^2 }{(p_1 - \Sigma \tau p_2)^2 + p_2^2} \exp \left[ -2 \int_{0}^{\tau} d \tau' \, \Gamma \left( \sqrt{(p_1 - \Sigma \tau' p_2)^2 + p_2^2} \right)  \right],&\\
&\displaystyle Q_{\nu} = 2 \nu \int_0^{\infty} d \tau \int \dfrac{d^2 \bm p \, p^2 \chi(\bm p) }{(2 \pi)^2} [(p_1 - \Sigma \tau p_2)^2 + p_2^2]^{q-1} \exp \left[ -2 \int_{0}^{\tau} d \tau' \, \Gamma \left( \sqrt{(p_1 - \Sigma \tau' p_2)^2 + p_2^2} \right)  \right].&
\end{eqnarray}
It can be shown that the first term is equal to $Q_{\alpha} = \alpha \langle u_r^2 + u_{\phi}^2 \rangle/\varepsilon$, and the energy of velocity fluctuations was analyzed in detail in Ref.~[15]. One must be careful, since this object is determined by the infrared integral. To demonstrate this explicitly, let us turn back to the wave vector $\bm k$, where $k_1 = p_1 - \Sigma \tau p_2$ and $k_2 = p_2$, and then
\begin{equation}
Q_{\alpha} = 2 \alpha \int \dfrac{d^2 \bm k}{(2 \pi)^2} \int_0^{\infty} d \tau  \dfrac{(k_1+\Sigma \tau k_2)^2 + k_2^2 }{k_1^2+k_2^2} \chi(k_1+\Sigma \tau k_2, k_2) \exp \left[ -2 \int_{0}^{\tau} d \tau' \, \Gamma \left( \sqrt{(k_1 + \Sigma \tau' k_2)^2 + k_2^2} \right)  \right].
\end{equation}
Next, we consider the small wave number $k \ll k_f$. The pumping covariance spectrum $\chi(\bm q) \simeq \frac{\sqrt{2\pi}}{\delta_f k_f} \exp \left( - (q-k_f)^2/2 \delta_f^2 \right)$ is non-zero in a ring of radius $k_f$ and width $\delta_f \ll k_f$, therefore the integrand over $\tau$ is non-zero only during the time interval $\Delta \tau \sim \delta_f/(|\Sigma| k)$, and we come to the estimate
\begin{equation}
Q_{\alpha} \sim \alpha \int \dfrac{k dk}{k^2} \dfrac{\delta_f}{|\Sigma| k} k_f^2 \dfrac{1}{\delta_f k_f} \exp \left(- \dfrac{k_f \Gamma(k_f)}{|\Sigma| k} \right) = \dfrac{\alpha k_f}{|\Sigma|} \int \dfrac{dk}{k^2} \exp \left(- \dfrac{k_f \Gamma(k_f)}{|\Sigma| k} \right).
\end{equation}
\end{widetext}
The presence of the exponent causes the integral to formally converge, but this happens on the wavenumbers $k < k_f \Gamma(k_f)/|\Sigma|$. For typical parameters $k_f \sim 10^2$, $\Gamma(k_f) \sim 10^{-4}$, $|\Sigma| \sim 1$, it can be found that the corresponding length scale is much larger than the vortex size, so the local approximation for velocity fluctuations is violated and the result of the formal calculation becomes incorrect. As a rough estimate, we can cut off the integral by the size $R \sim 1$ of the coherent vortex, and then we will get $Q_{\alpha} \sim 0.01$. Alternatively, we can use the expression $Q_{\alpha} = \alpha \langle u_r^2 + u_{\phi}^2 \rangle/\varepsilon$ and measure the value of $\langle u_r^2 + u_{\phi}^2 \rangle$ in DNS, see Fig.~3 in the main text. The second method leads to a similar estimate.

The term $Q_{\nu}$ should be calculated using numerical integration. For our parameters, the pumping covariance spectrum can be safely replaced by $\chi(\bm q) = 2 \pi \delta(q-k_f)/k_f$, since $\delta_f \ll k_f$, that simplifies the problem. We found that the remaining integrals are better computed using local adaptive methods. The obtained values are much larger than $Q_{\alpha}$ and for this reason $Q \simeq Q_{\nu}$. The corresponding expression was written in the main text, see equation~(5).

\section{Turbulent viscosity}\label{app:B}

To take into account the self-action of fluctuations, we propose to add an additional term $\nu_T k^2$ into $\Gamma(k)$, corresponding to eddy viscosity. Now $Q=Q_{\alpha}+Q_{\nu}+Q_T$, and the term $Q_{\alpha}$ can be neglected as before. For the remaining terms we found
\begin{widetext}
\begin{eqnarray}
&\displaystyle Q_{\nu} = 2 \nu \int_0^{\infty} d \tau \int \dfrac{d^2 \bm p \, p^2 \chi(\bm p) }{(2 \pi)^2} [(p_1 - \Sigma \tau p_2)^2 + p_2^2]^{q-1} \exp \left[ -2 \int_{0}^{\tau} d \tau' \, \Gamma \left( \sqrt{(p_1 - \Sigma \tau' p_2)^2 + p_2^2} \right)  \right],&\\
&\displaystyle Q_T = 2 \nu_T \int_0^{\infty} d \tau \int \dfrac{d^2 \bm p}{(2 \pi)^2} \chi(\bm p) p^2 \exp \left[ -2 \int_{0}^{\tau} d \tau' \, \Gamma \left( \sqrt{(p_1 - \Sigma \tau' p_2)^2 + p_2^2} \right)  \right],&
\end{eqnarray}
\end{widetext}
where $\Gamma (k) = \alpha + \nu k^{2q} + \nu_T k^2$. Again, for simplicity the pumping covariance spectrum can be safely replaced by $\chi(\bm q) = 2 \pi \delta(q-k_f)/k_f$, and the remaining integrals should be computed numerically. Note that the model for turbulent viscosity $\nu_T = const/|\Sigma(r)|$ used in the main text means that the value of $\nu_T$ is greater near the periphery of the coherent vortex, since $|\Sigma|$ decreases with increasing $r$. This dependence is significant, since for a typical vortex the value of $|\Sigma|$ changes up to $2$ orders of magnitude.


\begin{thebibliography}{31}%
\makeatletter
\providecommand \@ifxundefined [1]{%
 \@ifx{#1\undefined}
}%
\providecommand \@ifnum [1]{%
 \ifnum #1\expandafter \@firstoftwo
 \else \expandafter \@secondoftwo
 \fi
}%
\providecommand \@ifx [1]{%
 \ifx #1\expandafter \@firstoftwo
 \else \expandafter \@secondoftwo
 \fi
}%
\providecommand \natexlab [1]{#1}%
\providecommand \enquote  [1]{``#1''}%
\providecommand \bibnamefont  [1]{#1}%
\providecommand \bibfnamefont [1]{#1}%
\providecommand \citenamefont [1]{#1}%
\providecommand \href@noop [0]{\@secondoftwo}%
\providecommand \href [0]{\begingroup \@sanitize@url \@href}%
\providecommand \@href[1]{\@@startlink{#1}\@@href}%
\providecommand \@@href[1]{\endgroup#1\@@endlink}%
\providecommand \@sanitize@url [0]{\catcode `\\12\catcode `\$12\catcode
  `\&12\catcode `\#12\catcode `\^12\catcode `\_12\catcode `\%12\relax}%
\providecommand \@@startlink[1]{}%
\providecommand \@@endlink[0]{}%
\providecommand \url  [0]{\begingroup\@sanitize@url \@url }%
\providecommand \@url [1]{\endgroup\@href {#1}{\urlprefix }}%
\providecommand \urlprefix  [0]{URL }%
\providecommand \Eprint [0]{\href }%
\providecommand \doibase [0]{https://doi.org/}%
\providecommand \selectlanguage [0]{\@gobble}%
\providecommand \bibinfo  [0]{\@secondoftwo}%
\providecommand \bibfield  [0]{\@secondoftwo}%
\providecommand \translation [1]{[#1]}%
\providecommand \BibitemOpen [0]{}%
\providecommand \bibitemStop [0]{}%
\providecommand \bibitemNoStop [0]{.\EOS\space}%
\providecommand \EOS [0]{\spacefactor3000\relax}%
\providecommand \BibitemShut  [1]{\csname bibitem#1\endcsname}%
\let\auto@bib@innerbib\@empty
\bibitem [{\citenamefont {Kraichnan}(1967)}]{kraichnan1967inertial}%
  \BibitemOpen
  \bibfield  {author} {\bibinfo {author} {\bibfnamefont {R.~H.}\ \bibnamefont
  {Kraichnan}},\ }\href@noop {} {\bibfield  {journal} {\bibinfo  {journal} {The
  Physics of Fluids}\ }\textbf {\bibinfo {volume} {10}},\ \bibinfo {pages}
  {1417} (\bibinfo {year} {1967})}\BibitemShut {NoStop}%
\bibitem [{\citenamefont {Leith}(1968)}]{leith1968diffusion}%
  \BibitemOpen
  \bibfield  {author} {\bibinfo {author} {\bibfnamefont {C.~E.}\ \bibnamefont
  {Leith}},\ }\href@noop {} {\bibfield  {journal} {\bibinfo  {journal} {The
  Physics of Fluids}\ }\textbf {\bibinfo {volume} {11}},\ \bibinfo {pages}
  {671} (\bibinfo {year} {1968})}\BibitemShut {NoStop}%
\bibitem [{\citenamefont {Batchelor}(1969)}]{batchelor1969computation}%
  \BibitemOpen
  \bibfield  {author} {\bibinfo {author} {\bibfnamefont {G.~K.}\ \bibnamefont
  {Batchelor}},\ }\href@noop {} {\bibfield  {journal} {\bibinfo  {journal} {The
  Physics of Fluids}\ }\textbf {\bibinfo {volume} {12}},\ \bibinfo {pages} {II}
  (\bibinfo {year} {1969})}\BibitemShut {NoStop}%
\bibitem [{\citenamefont {Sommeria}(1986)}]{sommeria1986experimental}%
  \BibitemOpen
  \bibfield  {author} {\bibinfo {author} {\bibfnamefont {J.}~\bibnamefont
  {Sommeria}},\ }\href@noop {} {\bibfield  {journal} {\bibinfo  {journal}
  {Journal of Fluid Mechanics}\ }\textbf {\bibinfo {volume} {170}},\ \bibinfo
  {pages} {139} (\bibinfo {year} {1986})}\BibitemShut {NoStop}%
\bibitem [{\citenamefont {Paret}\ and\ \citenamefont
  {Tabeling}(1998)}]{paret1998intermittency}%
  \BibitemOpen
  \bibfield  {author} {\bibinfo {author} {\bibfnamefont {J.}~\bibnamefont
  {Paret}}\ and\ \bibinfo {author} {\bibfnamefont {P.}~\bibnamefont
  {Tabeling}},\ }\href@noop {} {\bibfield  {journal} {\bibinfo  {journal}
  {Physics of Fluids}\ }\textbf {\bibinfo {volume} {10}},\ \bibinfo {pages}
  {3126} (\bibinfo {year} {1998})}\BibitemShut {NoStop}%
\bibitem [{\citenamefont {Smith}\ and\ \citenamefont
  {Yakhot}(1993)}]{smith1993bose}%
  \BibitemOpen
  \bibfield  {author} {\bibinfo {author} {\bibfnamefont {L.~M.}\ \bibnamefont
  {Smith}}\ and\ \bibinfo {author} {\bibfnamefont {V.}~\bibnamefont {Yakhot}},\
  }\href@noop {} {\bibfield  {journal} {\bibinfo  {journal} {Physical Review
  Letters}\ }\textbf {\bibinfo {volume} {71}},\ \bibinfo {pages} {352}
  (\bibinfo {year} {1993})}\BibitemShut {NoStop}%
\bibitem [{\citenamefont {Smithr}\ and\ \citenamefont
  {Yakhot}(1994)}]{smithr1994finite}%
  \BibitemOpen
  \bibfield  {author} {\bibinfo {author} {\bibfnamefont {L.~M.}\ \bibnamefont
  {Smithr}}\ and\ \bibinfo {author} {\bibfnamefont {V.}~\bibnamefont
  {Yakhot}},\ }\href@noop {} {\bibfield  {journal} {\bibinfo  {journal}
  {Journal of Fluid Mechanics}\ }\textbf {\bibinfo {volume} {274}},\ \bibinfo
  {pages} {115} (\bibinfo {year} {1994})}\BibitemShut {NoStop}%
\bibitem [{\citenamefont {Borue}(1994)}]{borue1994inverse}%
  \BibitemOpen
  \bibfield  {author} {\bibinfo {author} {\bibfnamefont {V.}~\bibnamefont
  {Borue}},\ }\href@noop {} {\bibfield  {journal} {\bibinfo  {journal}
  {Physical Review Letters}\ }\textbf {\bibinfo {volume} {72}},\ \bibinfo
  {pages} {1475} (\bibinfo {year} {1994})}\BibitemShut {NoStop}%
\bibitem [{\citenamefont {Chertkov}\ \emph {et~al.}(2007)\citenamefont
  {Chertkov}, \citenamefont {Connaughton}, \citenamefont {Kolokolov},\ and\
  \citenamefont {Lebedev}}]{chertkov2007dynamics}%
  \BibitemOpen
  \bibfield  {author} {\bibinfo {author} {\bibfnamefont {M.}~\bibnamefont
  {Chertkov}}, \bibinfo {author} {\bibfnamefont {C.}~\bibnamefont
  {Connaughton}}, \bibinfo {author} {\bibfnamefont {I.}~\bibnamefont
  {Kolokolov}},\ and\ \bibinfo {author} {\bibfnamefont {V.}~\bibnamefont
  {Lebedev}},\ }\href@noop {} {\bibfield  {journal} {\bibinfo  {journal}
  {Physical Review Letters}\ }\textbf {\bibinfo {volume} {99}},\ \bibinfo
  {pages} {084501} (\bibinfo {year} {2007})}\BibitemShut {NoStop}%
\bibitem [{\citenamefont {Laurie}\ \emph {et~al.}(2014)\citenamefont {Laurie},
  \citenamefont {Boffetta}, \citenamefont {Falkovich}, \citenamefont
  {Kolokolov},\ and\ \citenamefont {Lebedev}}]{laurie2014universal}%
  \BibitemOpen
  \bibfield  {author} {\bibinfo {author} {\bibfnamefont {J.}~\bibnamefont
  {Laurie}}, \bibinfo {author} {\bibfnamefont {G.}~\bibnamefont {Boffetta}},
  \bibinfo {author} {\bibfnamefont {G.}~\bibnamefont {Falkovich}}, \bibinfo
  {author} {\bibfnamefont {I.}~\bibnamefont {Kolokolov}},\ and\ \bibinfo
  {author} {\bibfnamefont {V.}~\bibnamefont {Lebedev}},\ }\href@noop {}
  {\bibfield  {journal} {\bibinfo  {journal} {Physical Review Letters}\
  }\textbf {\bibinfo {volume} {113}},\ \bibinfo {pages} {254503} (\bibinfo
  {year} {2014})}\BibitemShut {NoStop}%
\bibitem [{\citenamefont {Frishman}\ and\ \citenamefont
  {Herbert}(2018)}]{frishman2018turbulence}%
  \BibitemOpen
  \bibfield  {author} {\bibinfo {author} {\bibfnamefont {A.}~\bibnamefont
  {Frishman}}\ and\ \bibinfo {author} {\bibfnamefont {C.}~\bibnamefont
  {Herbert}},\ }\href@noop {} {\bibfield  {journal} {\bibinfo  {journal}
  {Physical Review Letters}\ }\textbf {\bibinfo {volume} {120}},\ \bibinfo
  {pages} {204505} (\bibinfo {year} {2018})}\BibitemShut {NoStop}%
\bibitem [{\citenamefont {Xia}\ \emph {et~al.}(2009)\citenamefont {Xia},
  \citenamefont {Shats},\ and\ \citenamefont {Falkovich}}]{xia2009spectrally}%
  \BibitemOpen
  \bibfield  {author} {\bibinfo {author} {\bibfnamefont {H.}~\bibnamefont
  {Xia}}, \bibinfo {author} {\bibfnamefont {M.}~\bibnamefont {Shats}},\ and\
  \bibinfo {author} {\bibfnamefont {G.}~\bibnamefont {Falkovich}},\ }\href@noop
  {} {\bibfield  {journal} {\bibinfo  {journal} {Physics of Fluids}\ }\textbf
  {\bibinfo {volume} {21}},\ \bibinfo {pages} {125101} (\bibinfo {year}
  {2009})}\BibitemShut {NoStop}%
\bibitem [{\citenamefont {Orlov}\ \emph {et~al.}(2018)\citenamefont {Orlov},
  \citenamefont {Brazhnikov},\ and\ \citenamefont
  {Levchenko}}]{orlov2018large}%
  \BibitemOpen
  \bibfield  {author} {\bibinfo {author} {\bibfnamefont {A.~V.}\ \bibnamefont
  {Orlov}}, \bibinfo {author} {\bibfnamefont {M.~Y.}\ \bibnamefont
  {Brazhnikov}},\ and\ \bibinfo {author} {\bibfnamefont {A.~A.}\ \bibnamefont
  {Levchenko}},\ }\href@noop {} {\bibfield  {journal} {\bibinfo  {journal}
  {JETP Letters}\ }\textbf {\bibinfo {volume} {107}},\ \bibinfo {pages} {157}
  (\bibinfo {year} {2018})}\BibitemShut {NoStop}%
\bibitem [{\citenamefont {Kolokolov}\ and\ \citenamefont
  {Lebedev}(2016{\natexlab{a}})}]{kolokolov2016structure}%
  \BibitemOpen
  \bibfield  {author} {\bibinfo {author} {\bibfnamefont {I.}~\bibnamefont
  {Kolokolov}}\ and\ \bibinfo {author} {\bibfnamefont {V.}~\bibnamefont
  {Lebedev}},\ }\href@noop {} {\bibfield  {journal} {\bibinfo  {journal}
  {Physical Review E}\ }\textbf {\bibinfo {volume} {93}},\ \bibinfo {pages}
  {033104} (\bibinfo {year} {2016}{\natexlab{a}})}\BibitemShut {NoStop}%
\bibitem [{\citenamefont {Kolokolov}\ and\ \citenamefont
  {Lebedev}(2016{\natexlab{b}})}]{kolokolov2016velocity}%
  \BibitemOpen
  \bibfield  {author} {\bibinfo {author} {\bibfnamefont {I.}~\bibnamefont
  {Kolokolov}}\ and\ \bibinfo {author} {\bibfnamefont {V.}~\bibnamefont
  {Lebedev}},\ }\href@noop {} {\bibfield  {journal} {\bibinfo  {journal}
  {Journal of Fluid Mechanics}\ }\textbf {\bibinfo {volume} {809}} (\bibinfo
  {year} {2016}{\natexlab{b}})}\BibitemShut {NoStop}%
\bibitem [{\citenamefont {Frishman}(2017)}]{frishman2017culmination}%
  \BibitemOpen
  \bibfield  {author} {\bibinfo {author} {\bibfnamefont {A.}~\bibnamefont
  {Frishman}},\ }\href@noop {} {\bibfield  {journal} {\bibinfo  {journal}
  {Physics of Fluids}\ }\textbf {\bibinfo {volume} {29}},\ \bibinfo {pages}
  {125102} (\bibinfo {year} {2017})}\BibitemShut {NoStop}%
\bibitem [{\citenamefont {Frishman}\ \emph {et~al.}(2017)\citenamefont
  {Frishman}, \citenamefont {Laurie},\ and\ \citenamefont
  {Falkovich}}]{frishman2017jets}%
  \BibitemOpen
  \bibfield  {author} {\bibinfo {author} {\bibfnamefont {A.}~\bibnamefont
  {Frishman}}, \bibinfo {author} {\bibfnamefont {J.}~\bibnamefont {Laurie}},\
  and\ \bibinfo {author} {\bibfnamefont {G.}~\bibnamefont {Falkovich}},\
  }\href@noop {} {\bibfield  {journal} {\bibinfo  {journal} {Physical Review
  Fluids}\ }\textbf {\bibinfo {volume} {2}},\ \bibinfo {pages} {032602}
  (\bibinfo {year} {2017})}\BibitemShut {NoStop}%
\bibitem [{\citenamefont {Falkovich}(2016)}]{falkovich2016interaction}%
  \BibitemOpen
  \bibfield  {author} {\bibinfo {author} {\bibfnamefont {G.}~\bibnamefont
  {Falkovich}},\ }\href@noop {} {\bibfield  {journal} {\bibinfo  {journal}
  {Proceedings of the Royal Society A: Mathematical, Physical and Engineering
  Sciences}\ }\textbf {\bibinfo {volume} {472}},\ \bibinfo {pages} {20160287}
  (\bibinfo {year} {2016})}\BibitemShut {NoStop}%
\bibitem [{\citenamefont {Woillez}\ and\ \citenamefont
  {Bouchet}(2017)}]{woillez2017theoretical}%
  \BibitemOpen
  \bibfield  {author} {\bibinfo {author} {\bibfnamefont {E.}~\bibnamefont
  {Woillez}}\ and\ \bibinfo {author} {\bibfnamefont {F.}~\bibnamefont
  {Bouchet}},\ }\href@noop {} {\bibfield  {journal} {\bibinfo  {journal} {EPL
  (Europhysics Letters)}\ }\textbf {\bibinfo {volume} {118}},\ \bibinfo {pages}
  {54002} (\bibinfo {year} {2017})}\BibitemShut {NoStop}%
\bibitem [{\citenamefont {Woillez}\ and\ \citenamefont
  {Bouchet}(2019)}]{woillez2019barotropic}%
  \BibitemOpen
  \bibfield  {author} {\bibinfo {author} {\bibfnamefont {E.}~\bibnamefont
  {Woillez}}\ and\ \bibinfo {author} {\bibfnamefont {F.}~\bibnamefont
  {Bouchet}},\ }\href@noop {} {\bibfield  {journal} {\bibinfo  {journal}
  {Journal of Fluid Mechanics}\ }\textbf {\bibinfo {volume} {860}},\ \bibinfo
  {pages} {577} (\bibinfo {year} {2019})}\BibitemShut {NoStop}%
\bibitem [{\citenamefont {Boffetta}\ and\ \citenamefont
  {Ecke}(2012)}]{boffetta2012two}%
  \BibitemOpen
  \bibfield  {author} {\bibinfo {author} {\bibfnamefont {G.}~\bibnamefont
  {Boffetta}}\ and\ \bibinfo {author} {\bibfnamefont {R.~E.}\ \bibnamefont
  {Ecke}},\ }\href@noop {} {\bibfield  {journal} {\bibinfo  {journal} {Annual
  Review of Fluid Mechanics}\ }\textbf {\bibinfo {volume} {44}},\ \bibinfo
  {pages} {427} (\bibinfo {year} {2012})}\BibitemShut {NoStop}%
\bibitem [{\citenamefont {Constantinou}\ \emph {et~al.}(2021)\citenamefont
  {Constantinou}, \citenamefont {Wagner}, \citenamefont {Siegelman},
  \citenamefont {Pearson},\ and\ \citenamefont
  {Palóczy}}]{GeophysicalFlowsJOSS}%
  \BibitemOpen
  \bibfield  {author} {\bibinfo {author} {\bibfnamefont {N.~C.}\ \bibnamefont
  {Constantinou}}, \bibinfo {author} {\bibfnamefont {G.~L.}\ \bibnamefont
  {Wagner}}, \bibinfo {author} {\bibfnamefont {L.}~\bibnamefont {Siegelman}},
  \bibinfo {author} {\bibfnamefont {B.~C.}\ \bibnamefont {Pearson}},\ and\
  \bibinfo {author} {\bibfnamefont {A.}~\bibnamefont {Palóczy}},\ }\href@noop
  {} {\bibfield  {journal} {\bibinfo  {journal} {Journal of Open Source
  Software}\ }\textbf {\bibinfo {volume} {6}},\ \bibinfo {pages} {3053}
  (\bibinfo {year} {2021})}\BibitemShut {NoStop}%
\bibitem [{\citenamefont {{See Supplemental Material at [URL will be inserted
  by publisher] for video fragments of coherent vortex dipoles under uniform
  and spatially localized random forcing.}}()}]{SM}%
  \BibitemOpen
  \bibinfo {author} {\bibnamefont {{See Supplemental Material at [URL will be
  inserted by publisher] for video fragments of coherent vortex dipoles under
  uniform and spatially localized random forcing.}}}\BibitemShut {Stop}%
\bibitem [{\citenamefont {Chan}\ \emph {et~al.}(2012)\citenamefont {Chan},
  \citenamefont {Mitra},\ and\ \citenamefont {Brandenburg}}]{chan2012dynamics}%
  \BibitemOpen
\bibfield  {author} {  }\bibfield  {author} {\bibinfo {author} {\bibfnamefont
  {C.-k.}\ \bibnamefont {Chan}}, \bibinfo {author} {\bibfnamefont
  {D.}~\bibnamefont {Mitra}},\ and\ \bibinfo {author} {\bibfnamefont
  {A.}~\bibnamefont {Brandenburg}},\ }\href@noop {} {\bibfield  {journal}
  {\bibinfo  {journal} {Physical Review E}\ }\textbf {\bibinfo {volume} {85}},\
  \bibinfo {pages} {036315} (\bibinfo {year} {2012})}\BibitemShut {NoStop}%
\bibitem [{\citenamefont {Parfenyev}\ and\ \citenamefont
  {Vergeles}(2021)}]{parfenyev2021influence}%
  \BibitemOpen
  \bibfield  {author} {\bibinfo {author} {\bibfnamefont {V.~M.}\ \bibnamefont
  {Parfenyev}}\ and\ \bibinfo {author} {\bibfnamefont {S.~S.}\ \bibnamefont
  {Vergeles}},\ }\href@noop {} {\bibfield  {journal} {\bibinfo  {journal}
  {Physics of Fluids}\ }\textbf {\bibinfo {volume} {33}},\ \bibinfo {pages}
  {115128} (\bibinfo {year} {2021})}\BibitemShut {NoStop}%
\bibitem [{\citenamefont {Doludenko}\ \emph {et~al.}(2021)\citenamefont
  {Doludenko}, \citenamefont {Fortova}, \citenamefont {Kolokolov},\ and\
  \citenamefont {Lebedev}}]{doludenko2021coherent}%
  \BibitemOpen
  \bibfield  {author} {\bibinfo {author} {\bibfnamefont {A.}~\bibnamefont
  {Doludenko}}, \bibinfo {author} {\bibfnamefont {S.}~\bibnamefont {Fortova}},
  \bibinfo {author} {\bibfnamefont {I.}~\bibnamefont {Kolokolov}},\ and\
  \bibinfo {author} {\bibfnamefont {V.}~\bibnamefont {Lebedev}},\ }\href@noop
  {} {\bibfield  {journal} {\bibinfo  {journal} {Physics of Fluids}\ }\textbf
  {\bibinfo {volume} {33}},\ \bibinfo {pages} {011704} (\bibinfo {year}
  {2021})}\BibitemShut {NoStop}%
\bibitem [{\citenamefont {Kellay}\ and\ \citenamefont
  {Goldburg}(2002)}]{kellay2002two}%
  \BibitemOpen
  \bibfield  {author} {\bibinfo {author} {\bibfnamefont {H.}~\bibnamefont
  {Kellay}}\ and\ \bibinfo {author} {\bibfnamefont {W.~I.}\ \bibnamefont
  {Goldburg}},\ }\href@noop {} {\bibfield  {journal} {\bibinfo  {journal}
  {Reports on Progress in Physics}\ }\textbf {\bibinfo {volume} {65}},\
  \bibinfo {pages} {845} (\bibinfo {year} {2002})}\BibitemShut {NoStop}%
\bibitem [{\citenamefont {Parfenyev}\ \emph {et~al.}(2016)\citenamefont
  {Parfenyev}, \citenamefont {Vergeles},\ and\ \citenamefont
  {Lebedev}}]{parfenyev2016nonlinear}%
  \BibitemOpen
  \bibfield  {author} {\bibinfo {author} {\bibfnamefont {V.}~\bibnamefont
  {Parfenyev}}, \bibinfo {author} {\bibfnamefont {S.~S.}\ \bibnamefont
  {Vergeles}},\ and\ \bibinfo {author} {\bibfnamefont {V.~V.}\ \bibnamefont
  {Lebedev}},\ }\href@noop {} {\bibfield  {journal} {\bibinfo  {journal} {JETP
  letters}\ }\textbf {\bibinfo {volume} {103}},\ \bibinfo {pages} {201}
  (\bibinfo {year} {2016})}\BibitemShut {NoStop}%
\bibitem [{\citenamefont {Yablonskii}\ \emph {et~al.}(2017)\citenamefont
  {Yablonskii}, \citenamefont {Kurbatov},\ and\ \citenamefont
  {Parfenyev}}]{yablonskii2017acoustic}%
  \BibitemOpen
  \bibfield  {author} {\bibinfo {author} {\bibfnamefont {S.}~\bibnamefont
  {Yablonskii}}, \bibinfo {author} {\bibfnamefont {N.}~\bibnamefont
  {Kurbatov}},\ and\ \bibinfo {author} {\bibfnamefont {V.}~\bibnamefont
  {Parfenyev}},\ }\href@noop {} {\bibfield  {journal} {\bibinfo  {journal}
  {Physical Review E}\ }\textbf {\bibinfo {volume} {95}},\ \bibinfo {pages}
  {012707} (\bibinfo {year} {2017})}\BibitemShut {NoStop}%
\bibitem [{\citenamefont {Molenaar}\ \emph {et~al.}(2004)\citenamefont
  {Molenaar}, \citenamefont {Clercx},\ and\ \citenamefont
  {Van~Heijst}}]{molenaar2004angular}%
  \BibitemOpen
  \bibfield  {author} {\bibinfo {author} {\bibfnamefont {D.}~\bibnamefont
  {Molenaar}}, \bibinfo {author} {\bibfnamefont {H.}~\bibnamefont {Clercx}},\
  and\ \bibinfo {author} {\bibfnamefont {G.}~\bibnamefont {Van~Heijst}},\
  }\href@noop {} {\bibfield  {journal} {\bibinfo  {journal} {Physica D:
  Nonlinear Phenomena}\ }\textbf {\bibinfo {volume} {196}},\ \bibinfo {pages}
  {329} (\bibinfo {year} {2004})}\BibitemShut {NoStop}%
\bibitem [{\citenamefont {Gallet}\ and\ \citenamefont
  {Young}(2013)}]{gallet2013two}%
  \BibitemOpen
  \bibfield  {author} {\bibinfo {author} {\bibfnamefont {B.}~\bibnamefont
  {Gallet}}\ and\ \bibinfo {author} {\bibfnamefont {W.~R.}\ \bibnamefont
  {Young}},\ }\href@noop {} {\bibfield  {journal} {\bibinfo  {journal} {Journal
  of Fluid Mechanics}\ }\textbf {\bibinfo {volume} {715}},\ \bibinfo {pages}
  {359} (\bibinfo {year} {2013})}\BibitemShut {NoStop}%
\end{thebibliography}

%

\end{document}